\documentclass[preprint,12pt]{elsarticle}
\usepackage{amssymb}
\usepackage{amsmath}
\journal{Subatomic Particles and Cosmology}
\begin{document}
\begin{frontmatter}
\title{Prompt and non-prompt production of charm hadrons in proton-proton collisions at the Large Hadron Collider using machine learning}
\author{Raghunath Sahoo\textsuperscript{a,\ref{note0}}, Suraj Prasad\textsuperscript{a}, Neelkamal Mallick\textsuperscript{a}, \\ Kangkan Goswami\textsuperscript{a}, and Gagan B. Mohanty\textsuperscript{b}} 
\affiliation{organization={Department of Physics, Indian Institute of Technology Indore}, \\ 
            addressline={Simrol}, 
            city={Indore},
            postcode={453552}, 
            state={Madhya Pradesh},
            country={India}}
\affiliation{organization={Tata Institute of Fundamental Research},
            addressline={Homi Bhabha Road}, 
            city={Mumbai},
            postcode={400005},
            country={India}}
\begin{abstract}
In this contribution, we use machine learning (ML) based models to separate the prompt and non-prompt production of heavy flavour hadrons, such as $D^0$ and J/$\psi$, in proton-proton collisions at LHC energies. For this purpose, we use PYTHIA~8 to generate events, which provides a good qualitative agreement with experimental measurements of charm hadron production. The input features for the ML models are experimentally measurable. The prediction accuracy of the ML models used in this study reaches up to 99\%. The ML models can be useful in providing precise track-level identification, which is not possible in experiments with traditional methods. The contribution also discusses future applications of the ML models to understand the production of prompt and non-prompt heavy quark hadrons.

\end{abstract}
\end{frontmatter}
\footnotetext[1]{Email: Raghunath.Sahoo@cern.ch (Presenter)\label{note0}}

\section{Introduction}
In the era of precision measurements, the studies of heavy flavour hadron production are topical, which aids in the current understanding of Quantum Chromodynamics and serves as a probe for the thermalized medium formation in heavy-ion collisions. The lightest open charm meson, $D^{0}$, and vector meson charmonia, J/$\psi$, are well suited for the motivation. Prompt $D^{0}$ and J/$\psi$ constitute the direct production or feed-down from higher charm states, the study of which provides an undisputed ground to test different theoretical models. Further, the non-prompt production includes weak decay from beauty hadrons; thus, non-prompt $D^{0}$ and J/$\psi$ serve as indirect probes to study beauty hadron production. Since the beauty hadrons decay via weak interaction to the non-prompt charm hadrons, they usually are accompanied by larger values of $\vec{L}$ (distance between the primary interaction vertex and particle decay vertex), and distance of closest approach (DCA)~\cite{Goswami:2024xrx, Prasad:2023zdd}. We use these properties of $D^0$ and J/$\psi$ to segregate their topological productions using Machine Learning (ML).

\section{Methodology}

\begin{figure}
    \centering
    \includegraphics[width=0.46\linewidth]{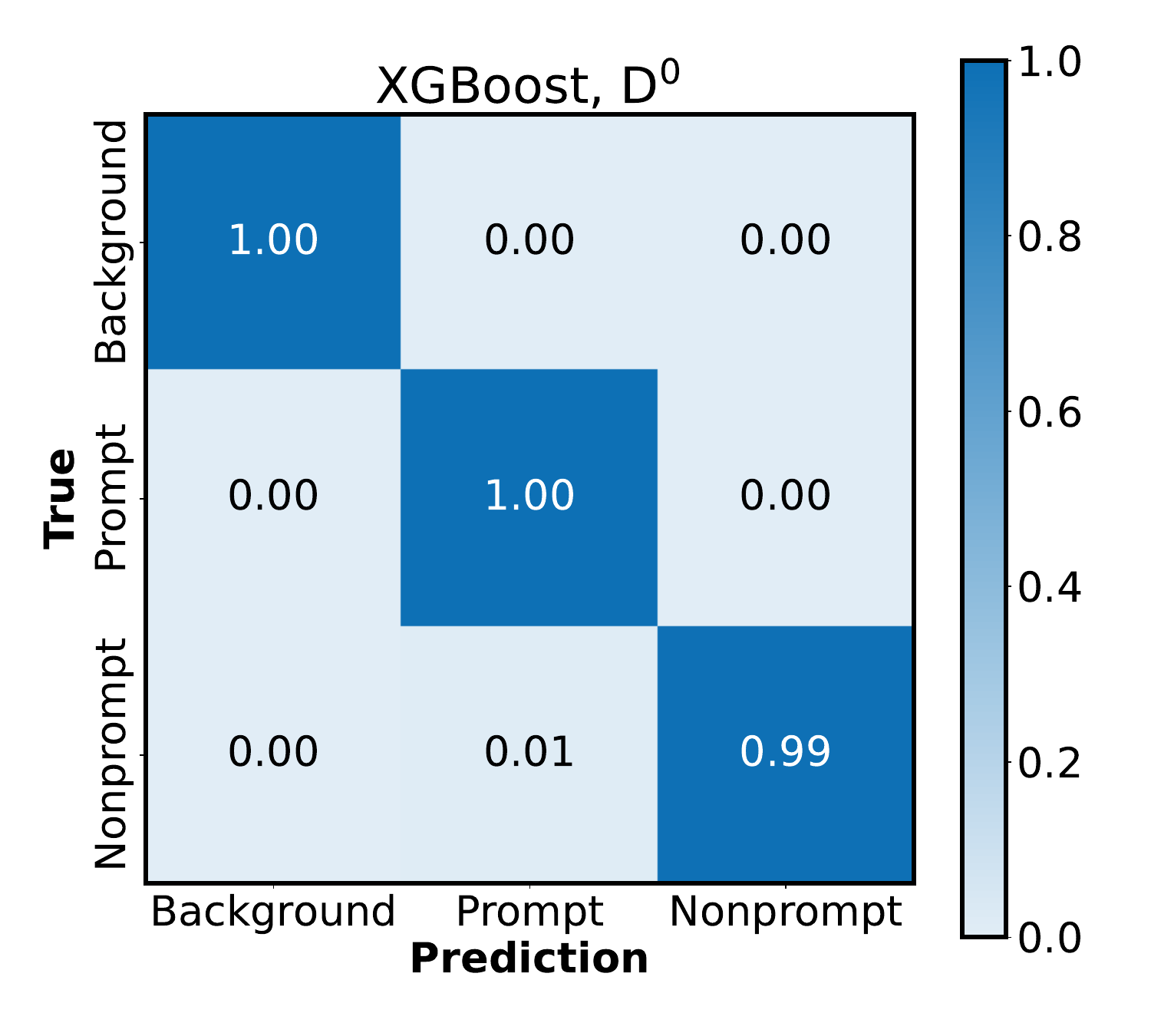}
    \includegraphics[width=0.46\linewidth]{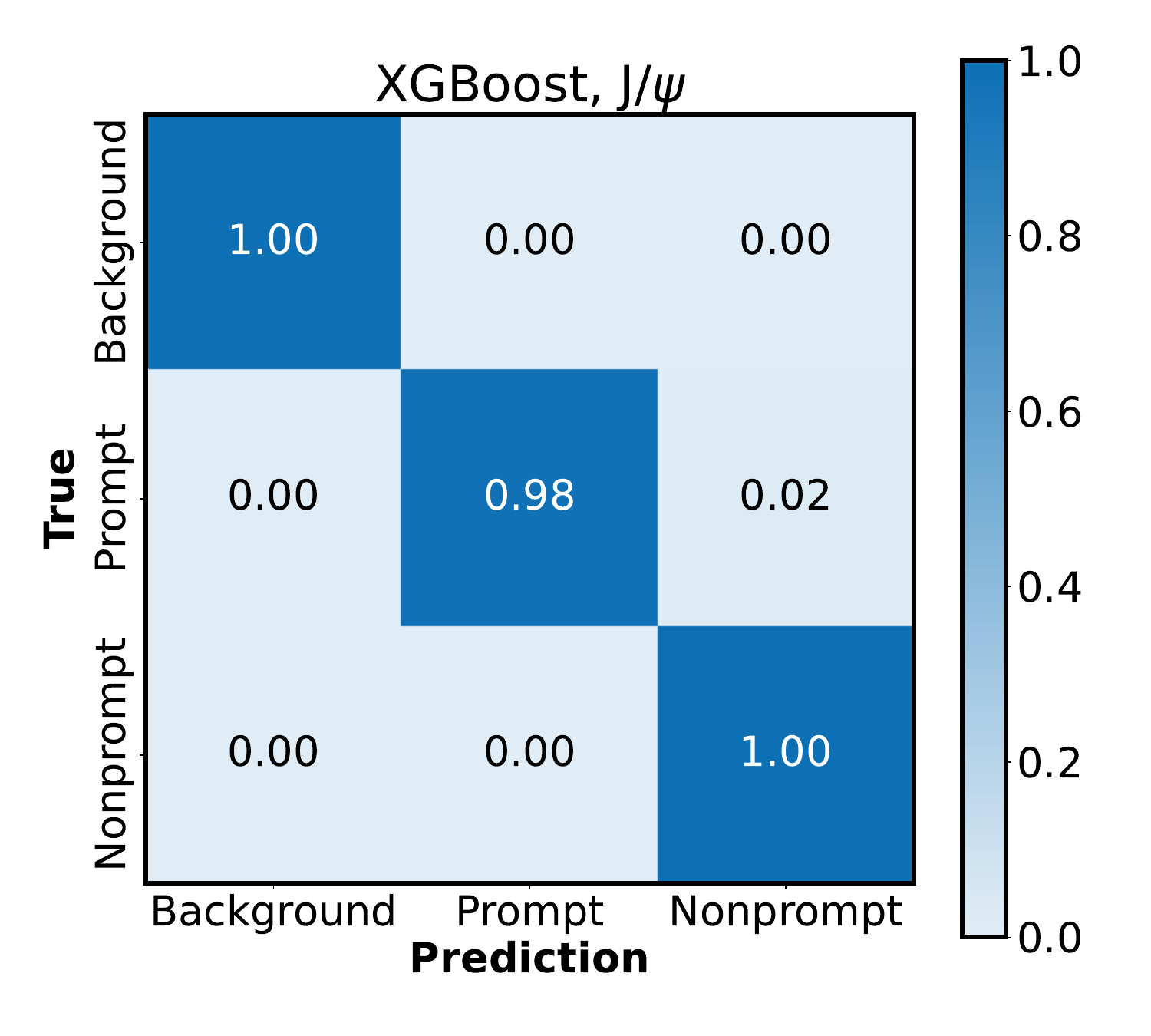}
    \caption{Confusion matrix of XGBoost model to separate prompt and non-prompt production of $D^0$ (left) and J/$\psi$ (right)~\cite{Goswami:2024xrx, Prasad:2023zdd}.}
    \label{fig:0}
\end{figure}

In this contribution, we use ML techniques to separate prompt and non-prompt production of $D^{0}$ and J/$\psi$ in pp collisions at the LHC energies. We use $D^{0}\rightarrow\pi^{+}K^{-}$ and J/$\psi\rightarrow\mu^{+}\mu^{-}$ channels for the study. For the prompt and non-prompt identifications of J/$\psi$, we use XGBoost (XGB) and LightGBM (LGBM) models. For $D^{0}$, we have considered Catboost, XGBoost and Random Forest models. The training dataset is created using the PYTHIA~8 Monte Carlo event generator. For topological segregation of J/$\psi$ mesons using ML, the input sample includes invariant mass ($m_{\mu\mu}$ ), transverse momentum, pseudorapidity, and $c{\tau}$. Similarly for $D^{0}$, we consider invariant mass ($m_{\pi K}$), pseudoproper time ($t_z$ ), pseudoproper decay length ($c{\tau}$ ), and distance of closest approach ($\rm{DCA}_{D_{0}}$) as the input features. A detailed discussion of PYTHIA~8 tuning for event generation, ML model parameters, and ML model performance can be found in Refs.~\cite{Goswami:2024xrx, Prasad:2023zdd}.

The confusion matrices are necessary to quantify and visualize the prediction accuracy for each class separately. Figure~\ref{fig:0} shows the confusion matrix for the XGBoost model, trained to separate the prompt and non-prompt production of $D^{0}$ (left panel) and J/$\psi$. For $D^0$, in the left panel, the XGBoost model completely separates the combinatorial background from the signal. However, 1\% of non-prompt $D^0$ is classified as prompt. Since the ratio of non-prompt to prompt is small, this 1\% misclassification does not affect the results, also shown in  Ref.~\cite{Goswami:2024xrx}. The prediction results are similar for CatBoost, while for Random Forest, this misclassification rises to 3\%. Thus, for $D^0$, we shall limit our results to XGB only. Similar to $D^0$ in the left panel, the classification accuracy for XGB to separate dimuon combinatorial background from signal J/$\psi$ is close to 100\%. However, 2\% of prompt J/$\psi$ is classified as non-prompt. A brief discussion on the consequence of this misclassification, as well as a fix, is discussed in Ref.~\cite{Prasad:2023zdd}.

\section{Results}
\begin{figure}
    \centering
    \includegraphics[width=0.46\linewidth]{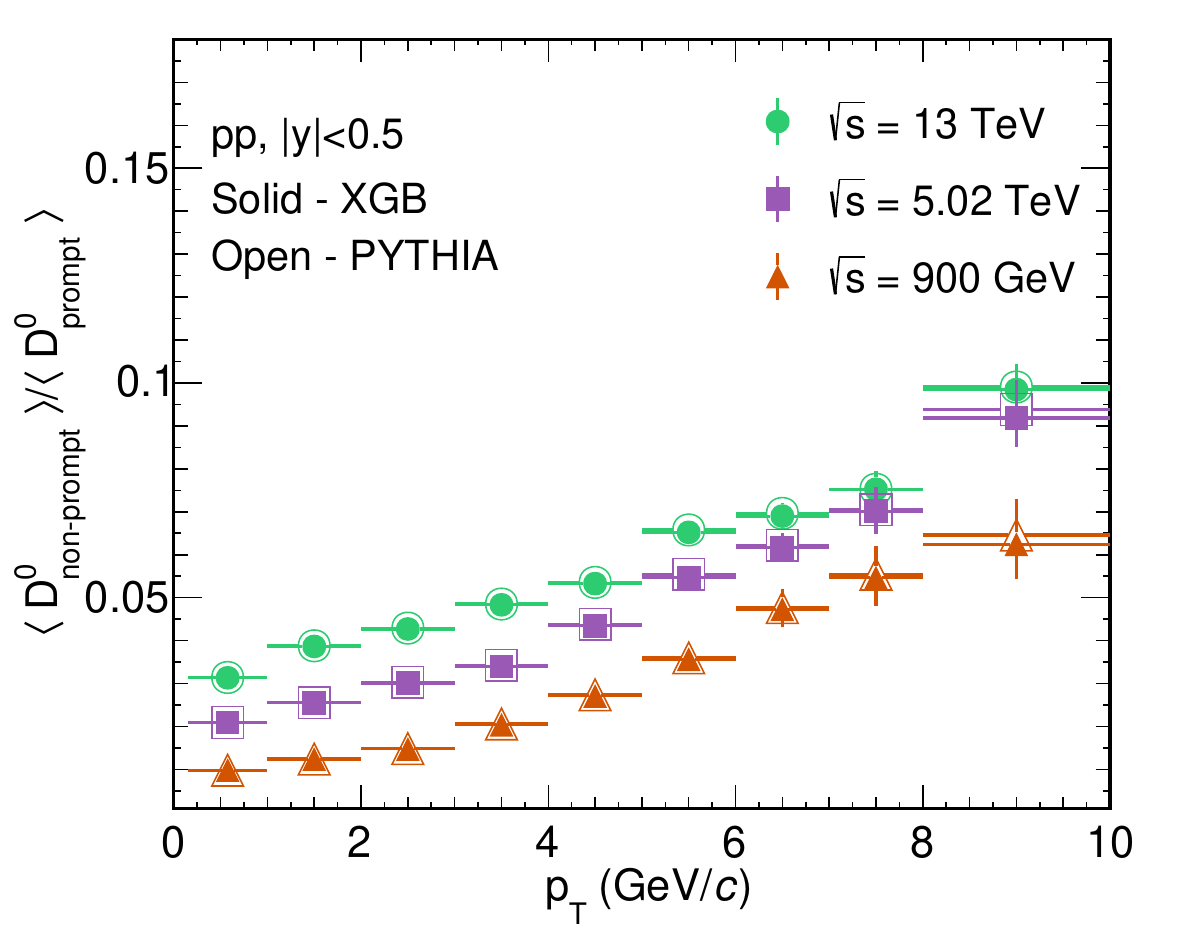}
    \includegraphics[width=0.46\linewidth]{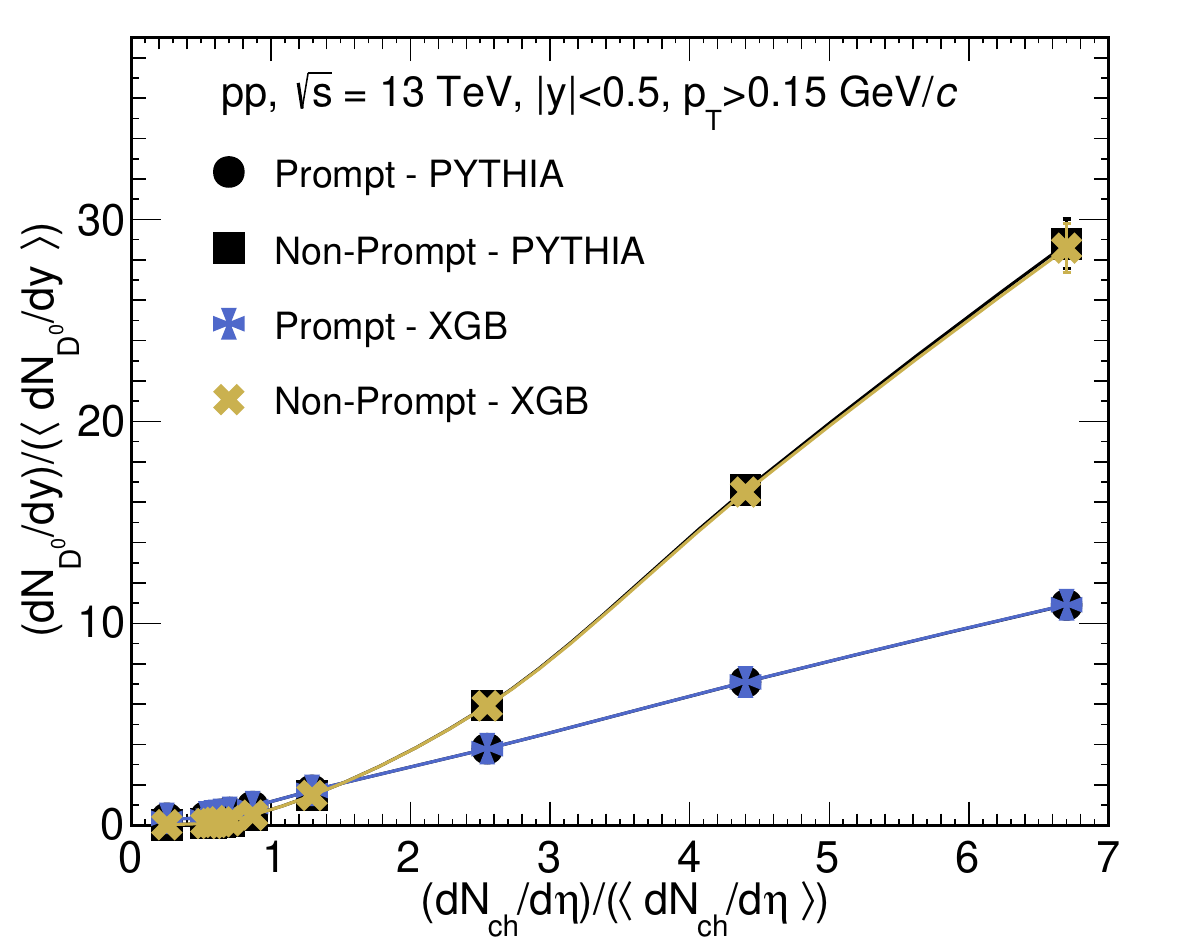}
    \caption{Left panel shows $p_{\rm T}$-differential non-prompt to prompt $D^0$ production fraction. The right panel shows the self-normalized yield ratio of $D^0$ as a function of self-normalized charged particle multiplicity in pp collisions using PYTHIA8 and XGB~\cite{Goswami:2024xrx}.}
    \label{fig:1}
\end{figure}

The left panel of Fig.~\ref{fig:1} shows $p_{\rm T}$-differential non-prompt to prompt $D^0$ production fraction ($\langle D^{0}_{\rm non-prompt}\rangle/\langle D^{0}_{\rm prompt}\rangle$) in pp collisions at $\sqrt{s}=13,\; 5.02,\;0.9$ TeV using PYTHIA~8 and XGB. Here $\langle D^{0}_{\rm non-prompt}\rangle/\langle D^{0}_{\rm prompt}\rangle$ increases with an increase in $p_{\rm T}$ and $\sqrt{s}$, which fairly indicates a rise in relative beauty hadron production with an increase in $p_{\rm T}$ and collision energy. A similar rise of non-prompt production fraction of J/$\psi$ is also observed in Ref.~\cite{Prasad:2023zdd}. The ML models provide accurate predictions for $\langle D^{0}_{\rm non-prompt}\rangle/\langle D^{0}_{\rm prompt}\rangle$. The right panel shows the self-normalized yield of prompt and non-prompt $D^0$ as a function of self-normalized charged particle density measured at mid-rapidity with event selection based on multiplicity at the V0 detector acceptance of ALICE~\cite{Goswami:2024xrx}. The self-normalized yield ratios show a stronger-than-linear increase with the increase in self-normalized charged particle density for both prompt and non-prompt cases. The effects are stronger for non-prompt cases. This is because when a beauty hadron is produced in initial hard scatterings, it is often accompanied by a jet in the opposite direction. These jet fragmentations tend to increase particle multiplicity, leading to a stronger-than-linear rise.
\begin{figure}
    \centering
    \includegraphics[width=0.46\linewidth]{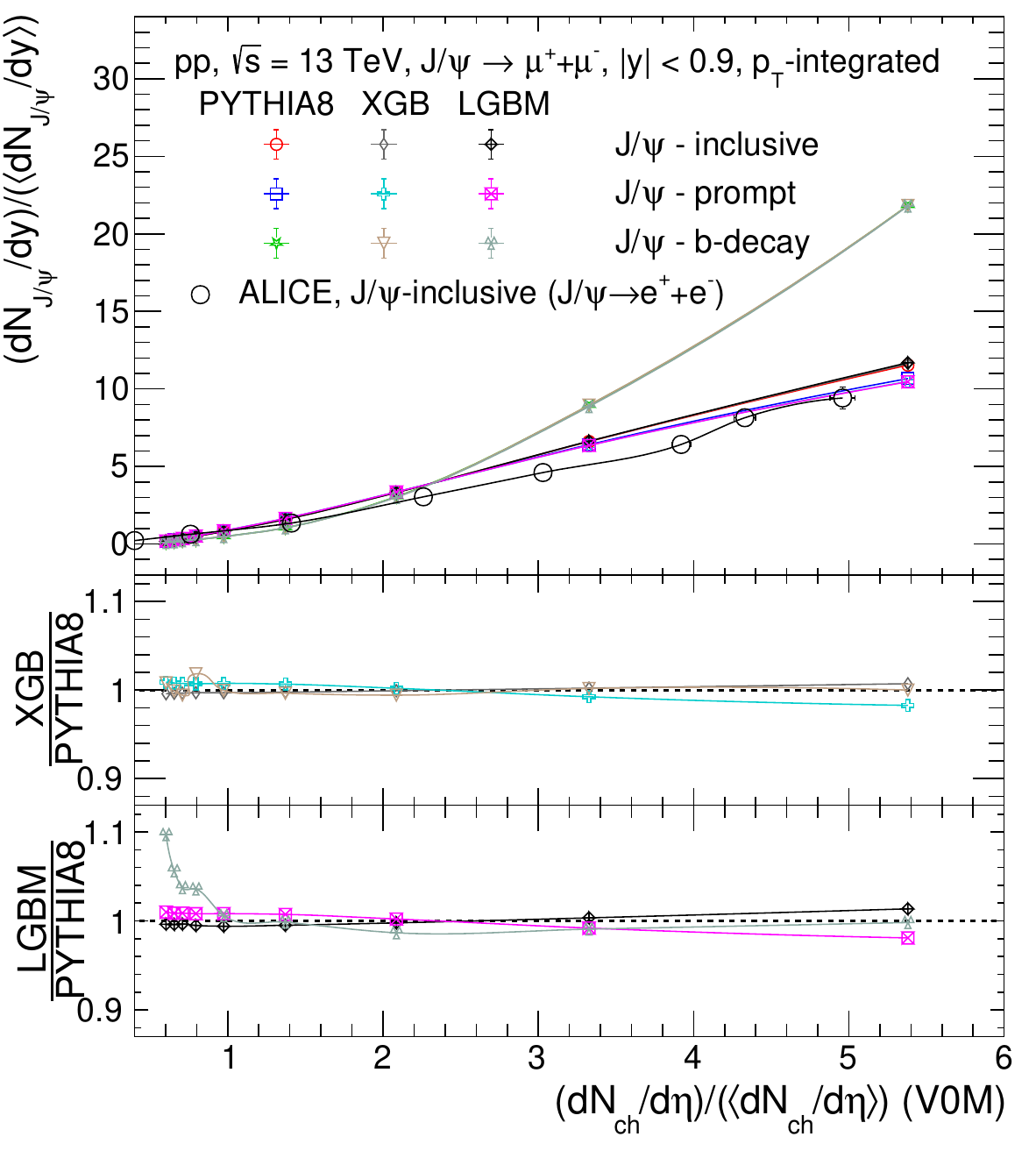}
    \includegraphics[width=0.46\linewidth]{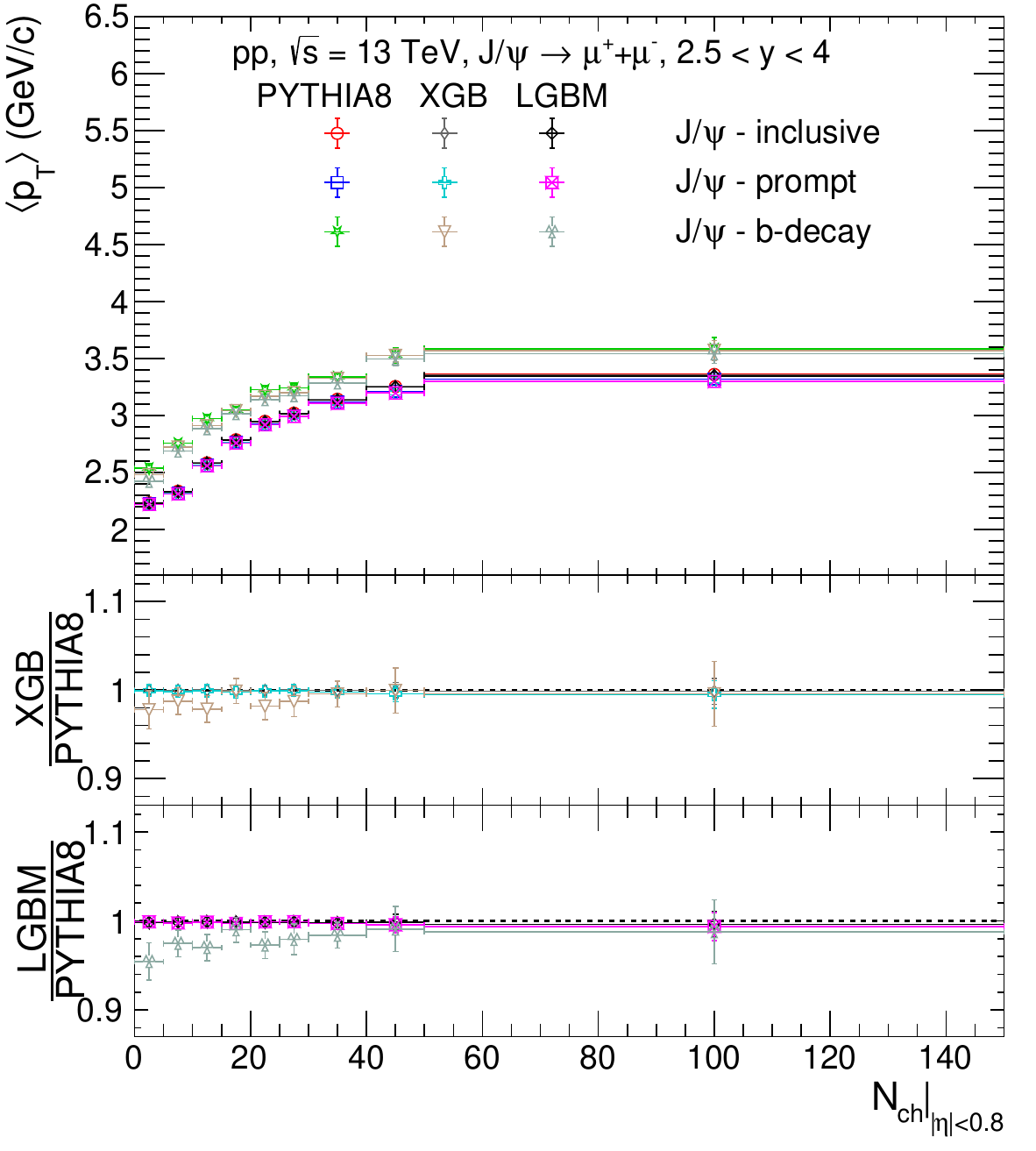}
    \caption{(Left panel) Self-normalized yield ratios of prompt and non-prompt ${\rm J}/\psi$ as a function of self-normalized charged particle multiplicity measured at mid-pseudorapidity and (right panel) $\langle p_{\rm T}\rangle$ of prompt and non-prompt ${\rm J}/\psi$ as a function of charged particle multiplicity in $|\eta|<0.8$ in pp collisions at $\sqrt{s}=13$ TeV using XGBoost, LightGBM and PYTHIA8~\cite{Prasad:2023zdd}. The ALICE measurements are taken from Ref.~\cite{ALICE:2021zkd}.}
    \label{fig:2}
\end{figure}

A similar stronger-than-linear rise of the self-normalized yield of prompt and non-prompt J/$\psi$ is observed, as shown in the left panel of Fig~\ref{fig:2}. However, for the prompt case, the stronger-than-linear is not trivially understood. One of the possible causes for these stronger-than-linear rises is speculated to arise due to biases from event selections, where different event shape observables come into play~\cite{Prasad:2024gqq}. To understand the production of prompt and non-prompt charm hadrons, we measure their average transverse momentum ($\langle p_{\rm T}\rangle$) as a function of charged particle multiplicity at mid-rapidity, shown in the right panel of Fig.~\ref{fig:2}. The non-prompt J/$\psi$ carries most of the momentum of the beauty-hadrons. Thus, their average momentum is higher than the prompt J/$\psi$. Further, a rise in $\langle p_{\rm T}\rangle$ of J/$\psi$ indicates an increase in opposite jet energy which leads to an increase in final state multiplicity. A study of prompt and non-prompt charm hadrons in different topological regions would be necessary to gain more insight into their production dynamics, where ML would be helpful in making precise measurements with track-level identifications.

\section{Summary}
In summary, we have exploited ML-based models to separate the prompt and non-prompt production of $D^0$ and J/$\psi$ in pp collisions. The classification accuracy of the discussed models reaches up to 99\%. The ML models successfully predicts the $p_{\rm T}$, $\eta$, $\sqrt{s}$ and multiplicity dependence of prompt and non-prompt $D^{0}$ and J/$\psi$~\cite{Goswami:2024xrx, Prasad:2023zdd}.
The stronger-than-linear rise of self-normalised heavy hadron production is well explained by PYTHIA~8, and ML predictions are in agreement with their true values. The ML-based studies provide more precise measurements with track-level identification than the traditional methods. This would not only reduce uncertainties but also help us to gain more insight into the production dynamics of charm hadrons.


\end{document}